\documentclass[twocolumn,rmp,nofootinbin]{revtex4}

\usepackage[final]{graphics}
\usepackage{amssymb}
\usepackage{amsmath}
\usepackage{latexsym}

\begin{document}


\title{Onset of DNA Aggregation in Presence of Monovalent 
and Multivalent Counterions
}
\thanks{\textit{Biophys. J.} 2003, in press.}
\author{Yoram Burak}
\email{yorambu@post.tau.ac.il}
\author{Gil Ariel}
\author{David Andelman}
\affiliation{School of Physics and Astronomy, \\
Raymond and Beverly Sackler Faculty of Exact Sciences \\
Tel Aviv University, Tel Aviv 69978, Israel}

\begin{abstract}
We address theoretically aggregation of DNA segments by
multivalent polyamines such as spermine and spermidine.
In experiments, the aggregation
occurs above a certain threshold concentration of
multivalent ions. We demonstrate that the dependence of this
threshold on the concentration of DNA has
a simple form. When the DNA concentration $c_{\rm DNA}$ is smaller than
the monovalent salt concentration, the threshold
multivalent ion concentration
depends linearly on $c_{\rm DNA}$,
having the form $\alpha c_{\rm DNA} + \beta$.
The coefficients $\alpha$ and $\beta$ are related to the density profile
of multivalent counterions around isolated DNA chains,
at the onset of their aggregation.
This analysis agrees extremely well with recent detailed
measurements on DNA aggregation in the presence of spermine. From
the fit to the experimental data, the number of 
condensed multivalent counterions per DNA chain can be deduced.
A few other conclusions can then be
reached: \textit{i}) the number of condensed spermine ions at the
onset of aggregation decreases with the addition of monovalent
salt; \textit{ii}) the Poisson-Boltzmann theory
over-estimates the number of condensed multivalent ions at high
monovalent salt concentrations; \textit{iii}) our analysis
of the data indicates that the DNA charge is not over-compensated by
spermine at the onset of aggregation.
\end{abstract}

\maketitle


\section*{INTRODUCTION}
Condensation and aggregation of DNA, induced by
multivalent counterions, have been extensively studied
in the past two decades
(for a review, see \textcite{Bloomfield_review} and references therein).
The term condensation usually refers to the collapse of
a single, long DNA chain. Condensation plays an important role
in storage and packing of DNA; for example, in viral capsids \cite{gelbart}.
Aggregation of DNA is a closely
related phenomenon, where multiple chains attract each other
and form a variety of condensed mesophases of complex structure \cite{Pelta,Pelta2}.
In both phenomena multivalent
counterions play a crucial role, screening the electrostatic
repulsion between charged strands of DNA and mediating an effective
attraction.

\begin{figure}
\scalebox{0.40}{\includegraphics{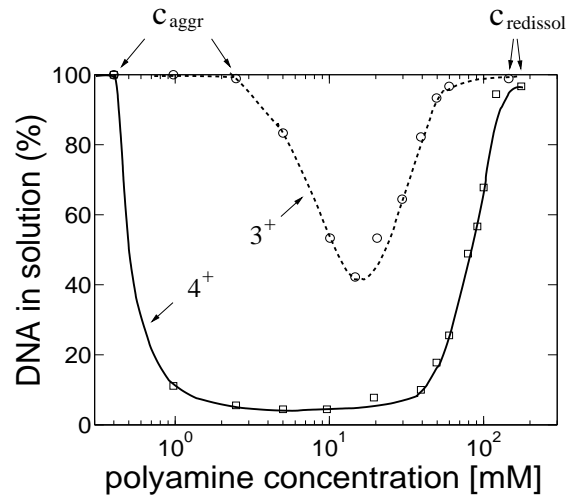}}
\caption{Percent of solubilized DNA, as
function of polyamine concentration: square symbols - spermine,
circles - spermidine. The solid and dashed lines are guides for
the eye. DNA and NaCl concentrations are $3\,{\rm mM}$ and
$25\,{\rm mM}$, respectively. Below the aggregation threshold,
$c_{\rm aggr}$, and above the re-dissolution threshold, $c_{\rm
redissol}$, all the DNA is dissolved. The data is adapted from
\textcite{Pelta}.
}
\end{figure}

A variety of tri- and tetra-valent ions can induce
aggregation and condensation, among them the polyamines
spermidine ($3^+$) and spermine ($4^+$)
\cite{Tabor,Gosule,Chattoraj}, as
well as cobalt-hexamine \cite{Widom1,Widom2}.
In typical experiments on aggregation
\cite{Pelta,Raspaud,Saminathan} multivalent ions are gradually
added to a solution with fixed concentration of DNA segments and 
monovalent salt. Two such examples for spermine and
spermidine are reproduced in Fig.~1 \cite{Pelta}. As the
multivalent ion concentration is raised above a certain threshold,
DNA segments begin to aggregate, and precipitate from the
solution.
Above the aggregation threshold, the
DNA concentration decreases gradually or abruptly,
depending on various parameters such as the monovalent salt
concentration and total DNA concentration. Further addition of
multivalent ions at higher concentrations reverses the
aggregation. Above a second, re-dissolution threshold,
all the DNA is re-dissolved in the solution (Fig.~1).
The re-dissolution threshold (above which all the DNA re-dissolves)
is almost independent on the DNA concentration. Its value can be
attributed to screening of electrostatic interactions by
multivalent ions \cite{Raspaud}.

The aggregation threshold, where the
onset of aggregation occurs, is the main experimental phenomenon
addressed in our theoretical paper. The multivalent ion
concentration at the onset depends strongly on the
monovalent salt and DNA concentrations. This dependence has
been recently measured in detail for short (150 base pair)
DNA segments in presence of spermine \cite{Raspaud}, and
is reproduced in Fig.~2. The figure shows measurements of spermine
concentrations at the onset of aggregation, for DNA concentrations
ranging over four orders of magnitude and for four different
monovalent salt concentrations: $2$, $13$, $23$ and $88\,{\rm
mM}$. At very low DNA concentration, the spermine concentration
depends strongly on the monovalent salt concentration. At higher
DNA concentration it has only a weak dependence on the monovalent
ion concentration but the spermine concentration is proportional to
the DNA concentration, indicating that a certain number of
spermine counterions are required, per DNA base, in order to
induce aggregation. The solid line in Fig.~2, adapted from
\textcite{Raspaud}, corresponds to a ratio: $c_{z,{\rm
aggr}}/c_{\rm DNA}=0.20$, where $c_{z,{\rm aggr}}$ 
is the spermine concentration at the aggregation 
onset and $c_{\rm DNA}$ is the DNA concentration. This linear relation
fits a large number of the
experimental points in the intermediate DNA concentration range.
It has been suggested by \textcite{Raspaud,Raspaud2} that the
deviations from this line, at low and high DNA concentrations,
represent two distinct physical regimes that need to be analyzed
separately from the intermediate regime, where the linear fit
works well.

\begin{figure}
\scalebox{0.45}{\includegraphics{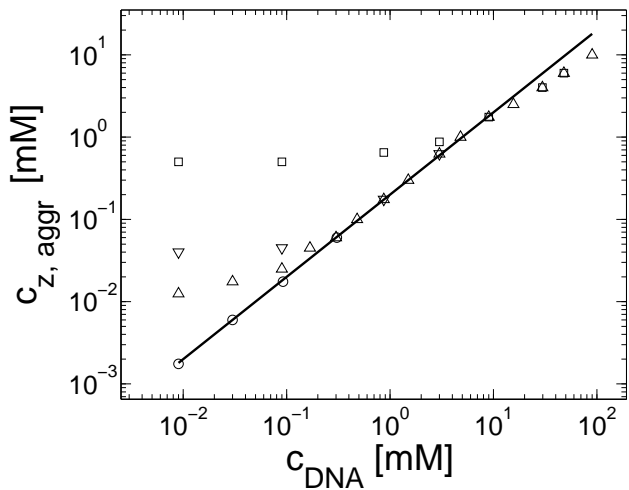}}
\caption{Spermine concentration $c_{z,{\rm
aggr}}$ at the onset aggregation, as a function of DNA monomer
concentration $c_{\rm DNA}$. Data is shown for four monovalent
salt concentrations:   $2\,{\rm mM}$ ($\circ$),   $13\,{\rm mM}$
($\triangle$),   $23\,{\rm mM}$ ($\triangledown$),  and $88\,{\rm
mM}$ ($\Box$). The solid line corresponds to a fixed ratio:
$c_{z,{\rm aggr}}/c_{\rm DNA}=0.20$. The data is adapted from
\textcite{Raspaud}.
}
\end{figure}

In this work we focus on the onset of aggregation, and
specifically on its dependence on the DNA concentration. We show
that this dependence is simple for {\it all the range} of DNA
concentration. Furthermore, for $c_{\rm DNA}$ smaller than the
monovalent salt concentration we show that this dependence is
linear: $c_{z,{\rm aggr}} = \alpha c_{\rm DNA} + \beta$. The
coefficient $\beta$ is the multivalent counterion concentration
far away from the DNA chains, while $\alpha$ accounts for the
excess of multivalent ions around each chain. These quantities can
be extracted, {\it e.g.,} from the four experimental curves of
Fig.~2. Several further conclusions are then drawn on the onset of
DNA aggregation and on the counterion distribution around each
double-stranded DNA.


\section*{THEORETICAL CONSIDERATIONS}

Consider an aqueous solution containing monovalent (1:1) salt,
multivalent ($z$:1) salt and DNA segments below their threshold
for aggregation. Throughout this paper, the DNA solution is
assumed to be dilute enough such that the DNA segments do not overlap.
We also assume that these DNA segments can be regarded as rigid rods.
The concentrations of
added monovalent salt, multivalent salt and DNA monomers are
denoted by  $c_s$, $c_z$ and $c_{\rm DNA}$,
respectively. 
These are the solute concentrations per unit volume 
as controlled and adjusted in experiments.
We will assume that the monovalent and multivalent salts
have the same type of co-ion, so that
altogether there are three ion species in the solution: 
\begin{enumerate}
\item A multivalent counterion contributed  
from the $z$:1 multivalent salt,
of concentration $c_z$.
\item A monovalent counterion contributed by
monovalent salt of concentration $c_s$, and by 
counterions dissociated from 
the DNA, of concentration $c_{\rm DNA}$: in total, 
$c_s+c_{\rm DNA}$.
\item Co-ions coming from both $z$:1 and 1:1 salts, of concentration
$c_s + z c_z$.
\end{enumerate}

Each DNA segment attracts a layer of oppositely charged
counterions referred to as the condensed counterions. As long as
the typical distance between segments is large compared to the
electrostatic screening length $\kappa^{-1}$, the electrostatic
potential decays exponentially to zero far away from the DNA segments.
In turn, the concentrations of the three ion species decay to well
defined {\it bulk} values denoted by $c^{\infty}_1$ for the
monovalent ions and $c^{\infty}_z$ for the $z$-valent ones.
These concentrations should be distinguished from the
concentrations $c_s$ and $c_z$ introduced above,
which are the average 
concentrations of added salts  regulated experimentally.

The Debye screening length, $\kappa^{-1}$, characterizing
the exponential decay of the electrostatic potential, is determined by the
bulk concentrations of all three ionic species:
\begin{equation}
\kappa^2 = 4\pi l_B\left[ c_1^{\infty} + z^2c_z^{\infty} + (c^{\infty}_1 + z c^{\infty}_z)\right].
\label{kappadef}
\end{equation}
where the third term is the co-ion concentration. It is equal
to $c^{\infty}_1 + z c^{\infty}_z$ due to charge neutrality far from the DNA where the
potential decays to zero.
The above equation makes use of the Bjerrum length,
$l_B = e^2/(\varepsilon k_B T)$, equal
to about $7$\,\AA~in aqueous solution at room temperature, $k_B T$
is the thermal energy , $e$ is the electron charge and $\varepsilon=80$
is the dielectric constant of water. The
Debye length as well as $c_z^\infty$ are shown schematically in
Fig.~3. Other quantities that will be defined below are also
indicated in this figure.

\begin{figure}
\scalebox{0.57}{\includegraphics*[5cm,2cm][21cm,11cm]{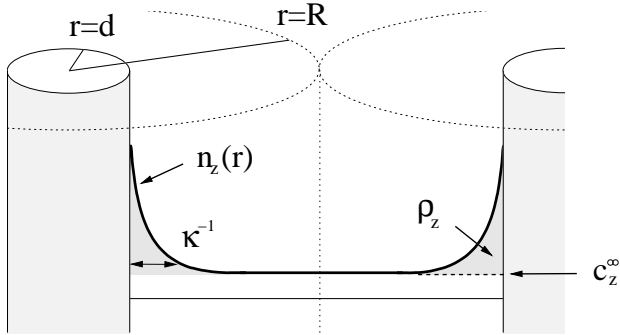}}
\caption{Schematic representation of the
multivalent density profile, $n_z(r)$ between two neighboring DNA
segments, each modeled as a cylinder of radius $d$.
Here $r$ is the distance from the axis of the left DNA
strand.
The radius $r=R$ corresponds to the inter-strand mid-distance
and is the unit cell radius.
The density decays to its bulk value $c_z^{\infty}$ on
distances larger than $\kappa^{-1}$, where $\kappa^{-1}$ is the
Debye length  defined in Eq.~\ref{kappadef}. The excess density of
multivalent ions $\rho_z$ is indicated by the shaded areas.
}
\end{figure}

In dilute solutions different DNA segments do not overlap.
Following previous works, we introduce a cell model also shown
schematically in Fig.~3. Note that the model serves
to illustrate the subsequent derivations but is not essential for
the validity of our main results. 
In the cell model, each segment,
of a cylindrical cross-section, is at the center of a cylindrical
cell of radius $R$ and area $A=\pi R^2$ such that
\begin{equation}
c_{\rm DNA} = 1/(aA) .
\label{ARdef}
\end{equation}
\noindent Namely, each DNA monomer occupies a specific volume
$aA$, where $a\simeq 1.7$\AA\ is the average charge separation
on the chain taken hereafter as the monomer length.

We will assume below that the DNA solution is dilute enough
so that $R$ is large compared to the Debye
length $\kappa^{-1}$. This assumption is essential for our derivation 
and can be verified for all the experimental data considered in this paper. 
Density profiles of the three ion species are then practically 
identical to those
near an isolated DNA segment with the same bulk concentrations
$c^{\infty}_1$, $c^{\infty}_z$. In other words, the profiles are
determined uniquely by $c^{\infty}_1$ and $c^{\infty}_z$, with
practically no dependence (or, more precisely, an exponentially
small dependence) on the DNA monomer concentration. A demonstration
of this claim is presented in Fig.~4, using the Poisson-Boltzmann
theory in a cell model. For two very different values of $R$
corresponding to different $c_{\rm DNA}$, the counterion profiles
match perfectly when the values of $c_1^\infty$ and
$c_z^\infty$ are the same. Note that the average
concentrations of added salts, $c_s$ and $c_z$, 
have different values in the two cells because of the contribution of 
condensed ions.

\begin{figure}
\scalebox{0.45}{\includegraphics{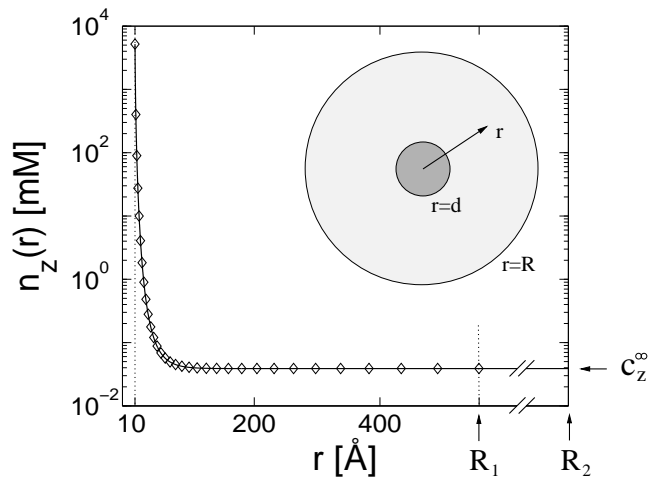}}
\caption{Density profile $n_z(r)$ of 4-valent
ions as function of $r$, the distance from the DNA axis, on a
semi-log plot, calculated using the
Poisson-Boltzmann equation in a cell model, where the
DNA segment
is modeled as a uniformly charged cylinder. The cell model is
shown schematically in the inset. Two cell sizes are shown, with
outer radii $R_1=560\,\mbox{\AA}$ ($c_{\rm DNA} = 1\,{\rm mM}$)
and $R_2=1.8 \times 10^4\,\mbox{\AA}$ ($c_{\rm DNA}=10^{-3}\,{\rm
mM}$), indicated by arrows.
In both cases the radius of closest approach of ions to the
charged chain is at $r=d$, where $d=10\,\mbox{\AA}$
as indicated by a dotted vertical line.
The boundary
condition at the inner cylinder matches the linear charge density
of DNA (1$e$/1.7\AA). The bulk densities of monovalent and
multivalent ions, $c_1^{\infty}$ and $c_z^{\infty}$,
are chosen to be the same in the two cells, leading to
practically identical density profiles. The solid line represents
the larger cell ($R_2$), and diamonds are used for the
smaller cell ($R_1$). Density profiles of monovalent counterions
and co-ions are not shown but are also practically identical in
the two cells. Average salt concentrations are $c_s=22\,{\rm mM}$ and
$c_z=0.21\,{\rm mM}$ in the smaller cell, and $c_s=23\,{\rm mM}$,
$c_z=0.039\,{\rm mM}$ in the larger cell. 
Bulk concentrations are $c_1^\infty=23$\,mM and 
$c_z^\infty = 0.039$\,mM. Note that these bulk concentrations
are practically identical to the salt concentrations in
the larger cell. Note also that $c_1^\infty > c_s$ in the smaller cell
reflecting the contribution of the counterions released by the DNA.
}
\end{figure}

The total number of $z$-valent counterions, per cell unit length,
is given by:
\begin{equation}
A c_z = A c^{\infty}_z + \rho_z(c^{\infty}_1,c^{\infty}_z)
\label{totalz}
\end{equation}
where $\rho_z$ is the excess number of $z$-valent ions per unit
length near the DNA. Throughout the paper we use the symbol $c$ to
denote concentrations per unit volume and $\rho$ for
concentrations per DNA unit length. The excess $\rho_z$
can be evaluated in the limit of infinite cell radius,
corresponding to an isolated chain:
\begin{equation}
\rho_z = 2 \pi \int_0^{\infty} r {\rm d}r\,\left[n_z(r)-c^{\infty}_z\right] ,
\label{excess}
\end{equation}
where $n_z(r)$ is the $z$-valent \textit{local} counterion
concentration at distance $r$ from the axis of symmetry, and
$n_z(\infty)=c_z^\infty$.
Following the discussion in
the previous paragraph, the excess $\rho_z$ is determined
uniquely by $c^{\infty}_1$ and $c^{\infty}_z$.
Its
exact functional dependence on these variables is generally
not known, although it can be evaluated approximately, \textit{e.g.}, 
using the Poisson-Boltzmann equation or in computer simulations.

For monovalent counterions we have, in a similar fashion:
\begin{equation}
A c_s + A c_{\rm DNA} = A c^{\infty}_1 + \rho_1(c^{\infty}_1,c^{\infty}_z) ,
\label{total1}
\end{equation}
where $\rho_1$, the excess of monovalent counterions per unit
length, is defined as in Eq.~\ref{excess}, and 
$Ac_{\rm DNA}=1/a$ is the DNA charge density per unit length.
The extra term in the left-hand-side of Eq.~\ref{total1} originates
from monovalent counterions contributed by the DNA monomers. Using
Eq.~\ref{ARdef} we can rewrite Eqs. \ref{totalz} and \ref{total1}
as:
\begin{equation}
c_z = c^{\infty}_z + a \rho_z\left(c^{\infty}_1,c^{\infty}_z\right) c_{\rm DNA}
\label{totalz_a}
\end{equation}
and
\begin{equation}
c_s = c^{\infty}_1 +
\left[a \rho_1\left(c^{\infty}_1,c^{\infty}_z\right) - 1\right]c_{\rm DNA}.
\label{total1_a}
\end{equation}
These two equations relate the experimentally adjustable $c_s$,
$c_z$ and $c_{\rm DNA}$ to the bulk densities $c_1^{\infty}$,
$c_z^{\infty}$ that in turn, are important because they determine
the ion density profiles.

In the limit of infinite DNA dilution, $c_{\rm DNA}= 0$, and
therefore $c_z = c^{\infty}_z$ and $c_s = c^{\infty}_1$. At any
finite DNA concentration $c_z$ and $c_s$ are not equal to
$c^{\infty}_z$ and $c^{\infty}_1$, respectively, because each
segment captures some of the multivalent ions and releases a
number of monovalent ones. Equations~\ref{totalz_a} and
\ref{total1_a} express the correction to $c_s, c_z$ at given
$c_1^\infty, c_z^\infty$ for both mono- and multi-valent
counterion species. The
dimensionless quantities $a\rho_1, a\rho_z$ are
the excess of the mono- and multi-valent counterion
species, respectively, per DNA monomer. 

We would like to emphasize  the generality of Eqs.
\ref{totalz_a} and \ref{total1_a}. They do not depend on
the assumption of parallel DNA residing in the middle of oriented cylindrical
unit cells, or on any mean-field approximation for the distribution of 
counterions. The only assumption required to derive  Eqs.
\ref{totalz_a} and \ref{total1_a} is
that the average distance between DNA segments 
is large compared with the Debye length. 
Although Eqs.~\ref{totalz_a} and \ref{total1_a} are correct for any $c_s$,
$c_z$ and $c_{\rm DNA}$ below the onset of DNA aggregation, we will be
interested below specifically in the aggregation onset.

\vspace*{1cm}
\noindent \textbf{Onset of aggregation}~~~\\

Our aim now is to find how the value
of $c_z$ at the onset of aggregation, $c_{z,{\rm aggr}}$,
depends on $c_{\rm DNA}$.
We will assume that this aggregation onset depends
on $c_1^{\infty}$ and $c_z^{\infty}$,
but not on the average distance between DNA chains.
We motivate this assumption by 
the fact that $c_1^\infty$ and
$c_z^\infty$ determine the density profile of multivalent 
counterions around the DNA chains, which, in turn, 
mediate the attraction necessary for aggregation. 
Before discussing this assumption in more detail,
let us first consider its consequences.
We can imagine an experiment where $c_z^{\infty}$
is gradually
increased while $c_1^{\infty}$ is kept fixed.
Aggregation will start, in this experiment,
above a certain threshold
value of $c_z^{\infty}$. Our assumption is that
this threshold does not depend on $c_{\rm DNA}$.
In real experiments, however, $c_z$ is adjusted rather
than $c_z^{\infty}$, and $c_s$ is kept fixed
rather than $c_1^{\infty}$.
In order to find the threshold value in terms of the
experimentally available
$c_z$ we need to map
$c_1^{\infty}, c_z^{\infty}$ onto $c_s, c_z$. This mapping is
described by Eqs.~\ref{totalz_a}--\ref{total1_a},
and involves $c_{\rm DNA}$. It is only through this mapping
that $c_{\rm DNA}$ will affect the threshold of aggregation.

\begin{figure*}
\scalebox{0.45}{\includegraphics{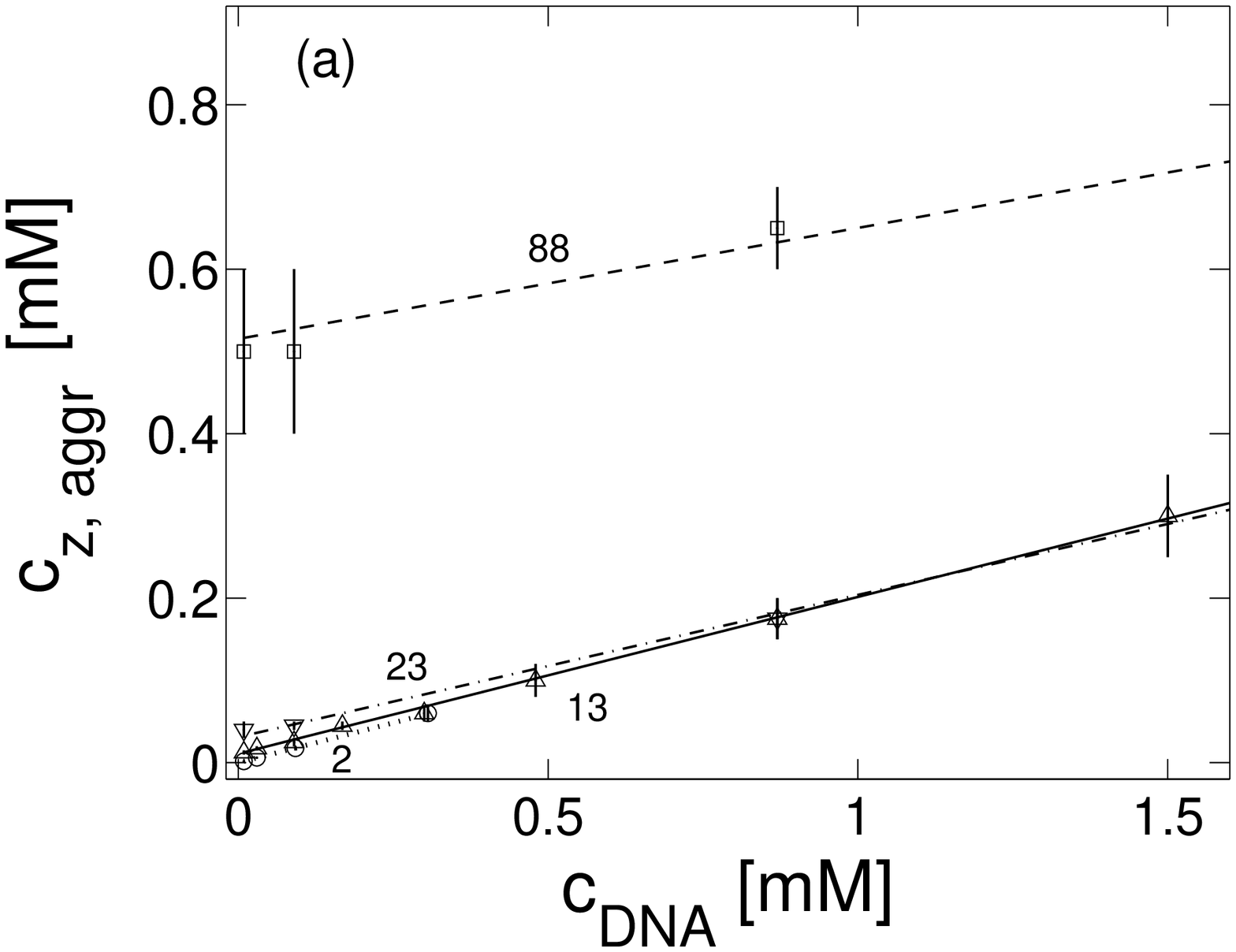}}
\hspace{0.7cm}
\scalebox{0.45}{\includegraphics{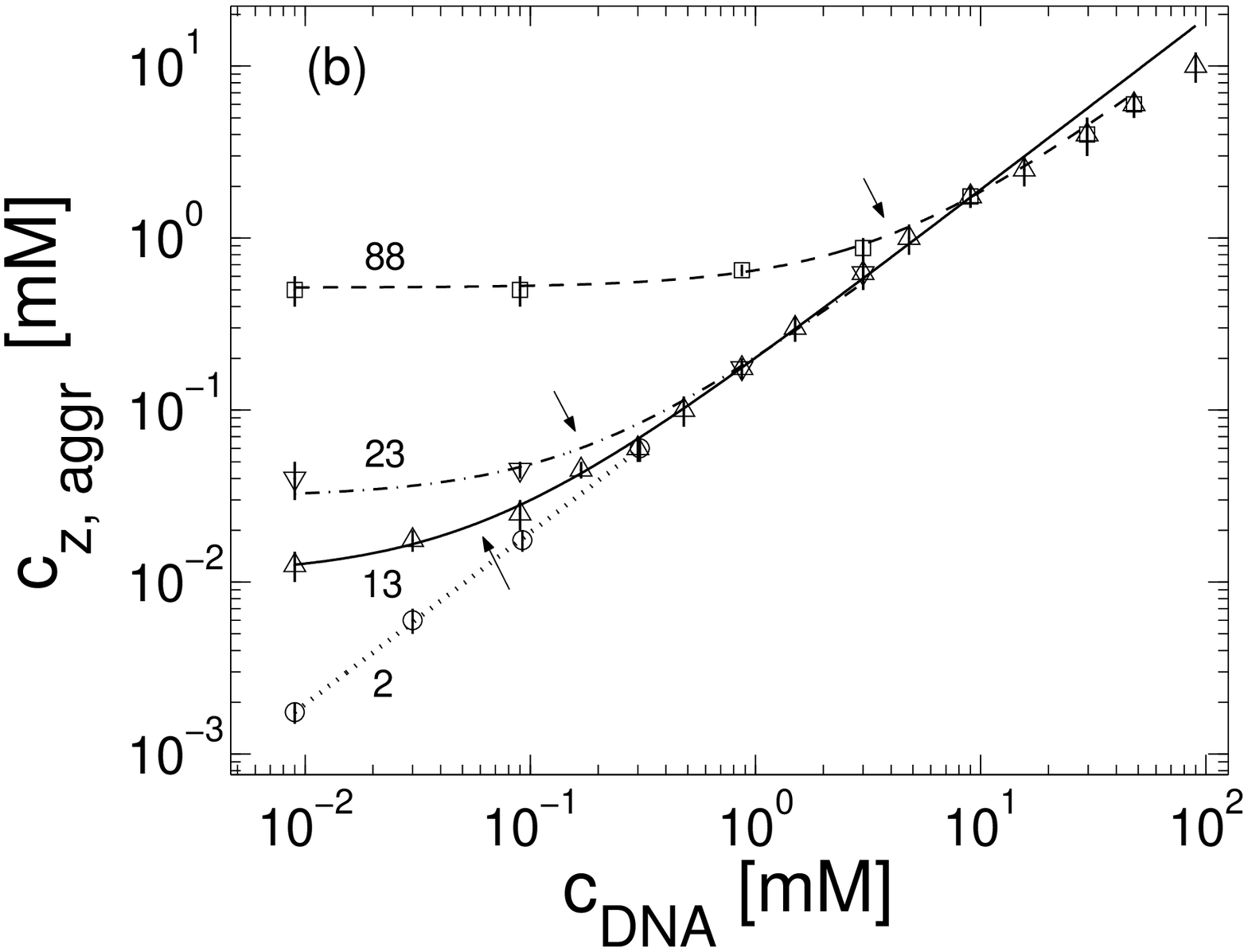}}
\caption{Spermine concentration at the
onset of aggregation $c_{z,{\rm aggr}}$
as a function of $c_{\rm DNA}$, fitted to the
form derived in Eq.~\ref{onset} (different line types are used for different salt
concentrations). Value of $c_s$
(in mM) is indicated next to each curve. Experimental data is adapted
from \textcite{Raspaud} and shown in the following symbols: $c_s =
2\,{\rm mM}$ ($\circ$),  $13\,{\rm mM}$ ($\triangle$), $23\,{\rm
mM}$ ($\triangledown$), and  $88\,{\rm mM}$ ($\Box$). Experimental
error bars (E. Raspaud, private communication)
are indicated by vertical lines. The fitted lines and
experimental points are shown using a linear scale in (a) up to
$c_{\rm DNA}=1.5$\,mM, and a log-log scale in (b)
up to $c_{\rm DNA}=100$\,mM, allowing all data
points to be shown on the same plot. Only the data up to $c_{\rm
DNA} = 10\,{\rm mM}$ was used for the linear fit. The crossover values of
$c_{\rm DNA}$, as defined by Eq.~\ref{crossover}, are indicated by
arrows in (b).
}
\end{figure*}

\vspace*{1cm}
\noindent \textit{The limit of $c_{\rm DNA} \ll c_s$}:~~~\\

The limit $c_{\rm DNA} \ll c_s$ offers a particularly simple
dependence of $c_{z,{\rm aggr}}$ on $c_{\rm DNA}$ and is considered first.
Most models and
experiments indicate that monovalent counterions cannot overcharge
DNA segments. Hence the monovalent excess, $a\rho_1$,
in Eq.~\ref{total1_a}, is a number
between zero and one, because the excess monovalent charge is
smaller than that of DNA. From Eq.~\ref{total1_a}
$|c_s - c^{\infty}_1|\ll c_s$ as long as $c_{\rm DNA} \ll c_s$.
It is then possible to
replace $c^{\infty}_1$ by $c_s$, leading
to a simplification of Eq.~\ref{totalz_a}:
\begin{equation}
c_z = c^{\infty}_z + a \rho_z\left(c_s,c^{\infty}_z\right) c_{\rm
DNA}. \label{totalz_b}
\end{equation}
Note that $c_{\rm DNA}$
is indeed smaller than $c_s$ in most of the experimental points in
Fig.~2. However a similar simplification cannot be applied for
$c_z$ because it is typically much smaller than $c_s$, and
often smaller than $c_{\rm DNA}$.

According to our principal assumption, aggregation
starts at a threshold value $c_z^{\infty} = c_z^*$, which does
not depend on $c_{\rm DNA}$ (while $c_{z,{\rm aggr}}$, the
average multivalent salt concentration does depend on $c_{\rm DNA}$
through Eq.~\ref{totalz_b}).
Similarly, the density profile at the
threshold does not depend
on $c_{\rm DNA}$, because it is determined by
$c_1^{\infty}=c_s$ and $c_z^*$. The excess of
$z$-valent counterions, as determined from this profile, is equal to:
\begin{equation}
\rho_z^* = \rho_z\left(c_s,c_z^*\right),
\end{equation}
\noindent with no dependence on $c_{\rm DNA}$.
Using the threshold values $c_z^*$ and $\rho_z^*$ in
Eq.~\ref{totalz_b}, we find that the average concentration of
$z$-valent ions at the onset of aggregation is:
\begin{equation}
c_{z,{\rm aggr}}(c_{\rm DNA}) = c_z^* + a \rho_z^* c_{\rm DNA}.
\label{onset}
\end{equation}
This is the threshold concentration that was measured
experimentally in \textcite{Raspaud}. Note that in Eq.~\ref{onset}
$c_z^*$ as well as $\rho_z^*$ depend on the monovalent salt
concentration, $c_s$, but the explicit
dependence is omitted for clarity.

The simple relationship expressed by Eq.~\ref{onset} is one of our
main results. As a visualization of this result we refer again to
Fig.~3. The quantities $\rho_z$, $c_z^{\infty}$ and the density
profile $n_z(r)$ are indicated in this figure. At the onset of
aggregation $c_z^{\infty}$ is equal to $c_z^*$ and does not depend
on $c_{\rm DNA}$ (or equivalently, on the spacing between DNA
segments, $R$). As $c_{\rm DNA}$ is increased the distance between
DNA strands decreases. The onset values of  $c_z^{\infty}$ and
$\rho_z$ do not change, but the contribution of $\rho_z$ to the
average concentration increases, leading to an increase in
$c_{z,\rm aggr}$.

The coefficients $a \rho_z^*$ and $c_z^*$ of the linear dependence
in Eq.~\ref{onset} are the coefficients $\alpha$ and $\beta$
defined in the introduction section. They
can be easily found from the experimental data:
$c_z^*$ is the value of $c_{z,\rm{aggr}}$ in the limit of infinite
DNA dilution, $c_{\rm DNA}\rightarrow 0$, since in this limit
$c_z=c_z^{\infty}=c_z^*$. The excess at the onset, $\rho_z^*$, can
be found from the slope of $c_{z,{\rm aggr}}$ as function of
$c_{\rm DNA}$. Before presenting a detailed comparison with experiments,
we generalize the treatment for small $c_{\rm DNA}$ to
arbitrary values.

\vspace*{1cm}
\noindent \textit{The case of $c_{\rm DNA}\ge c_s $ }:~~~\\

When $c_{\rm DNA}$ is of the same order as $c_s$ or larger,
corrections to $c^{\infty}_1$ must be taken into account, as
expressed by Eq.~\ref{total1_a}, and the linear relation
of Eq.~\ref{onset} no longer holds. The ion density profiles as well
as $c_s$ and $c_z$ are now determined by the two variables
$c_1^{\infty}$ and $c_z^{\infty}$. The relation between
$c_1^{\infty}$ and $c_z^{\infty}$ 
and the
experimentally controlled $c_s, c_z, c_{\rm DNA}$ is given by
Eqs.~\ref{totalz_a}-\ref{total1_a}. In terms of $c_1^{\infty},
c_z^{\infty}$ the criterion for aggregation remains the same as in
the previous case:
\begin{equation}
c^{\infty}_z = c_z^*(c^{\infty}_1).
\label{criterion}
\end{equation}
\noindent The three equations \ref{totalz_a}, \ref{total1_a} and
\ref{criterion}, with the three unknowns $c^{\infty}_1$,
$c^{\infty}_z$ and $c_z$ lead to a unique solution for $c_{z,{\rm
aggr}}$. Note that $c_1^{\infty}$ is
larger than $c_s$ because of counterions coming from
the DNA as can be seen in Eq.~\ref{total1_a}, where $a\rho_1-1$ is negative. In
Eq.~\ref{onset}, $c_s$ is replaced by $c_1^{\infty}$, which is
larger than $c_s$ for large $c_{\rm DNA}$. Hence, increasing
$c_{\rm DNA}$ has an effect similar to addition of monovalent
salt. As noted above, this effect is significant only for $c_{\rm
DNA} > c_s$.


\section*{COMPARISON WITH EXPERIMENT}

\textcite{Raspaud} measured the spermine ($z=4$) concentration
$c_z$ at the onset of aggregation as a function of $c_{\rm DNA}$
for four values of $c_s$ and with $c_{\rm DNA}$ ranging over four
orders of magnitude
--- from $10^{-2}$ to $10^2\,{\rm mM}$. We fitted the data (E.
Raspaud and J.-L. Sikorav, private communication) for each $c_s$
to a straight line according to Eq.~\ref{onset}. The least square
fit presented in Fig.~5 takes into account the experimental error
bars and data points up to $c_{\rm DNA}=10\,{\rm mM}$.
In Fig.~5\,\textit{a} the fit is shown using a linear scale which
covers the range of $c_{\rm DNA}$ only
up to $c_{\rm DNA}=1.5$\,mM for clarity purposes. Due to the large
range of $c_{\rm DNA}$ it is impossible to show all the data on the
linear scale of Fig.~5\,\textit{a}.
Instead, the same data and linear lines are shown
in Fig.~5\,\textit{b} on a log-log
scale over the full experimental range of $c_{\rm DNA}$.

The linear fit is very good for all four values of monovalent salt
concentration $c_s$. Note that for $c_s=88\,{\rm mM}$ the fit is
very good up to the largest value of $c_{\rm DNA}=48$\,mM reported
in the experiment, although our fit takes into account only data
points up to $c_{\rm DNA}=10$\,mM. It was previously suggested
\cite{Raspaud} that a separate regime exists for $c_{\rm DNA}
\gtrsim 10\,{\rm mM}$, characterized by a power law relation
between $c_z$ and $c_{\rm DNA}$ with an exponent smaller than
unity.
Our analysis suggests a different conclusion. The fit clearly
demonstrates that the relation is linear all the way up to $c_{\rm
DNA}=48$\,mM, as predicted by Eq.~\ref{onset}. Note also
that even at $c_{\rm DNA}=48$\,mM we have $c_{\rm DNA} < c_s$
so the assumptions leading to Eq.~\ref{onset} are still valid.

The only points in Fig.~5\,\textit{b} that deviate significantly
from the fit are the three points where $c_s=13\,{\rm mM}$
(triangles) and $c_{\rm DNA} > 20\,{\rm mM}$ (two of these points
coincide with points having $c_s=88\,{\rm mM}$, shown using square
symbols.) This deviation is easily explained by the fact that
$c_{\rm DNA} \gg c_s$ so that corrections to $c^{\infty}_1$ must
be taken into account. For example, at $c_{\rm DNA}=90\,{\rm mM}$
the nominal monovalent counterion concentration is $103\,{\rm
mM}$, taking into account counterions contributed by the DNA. In
order to find $c^{\infty}_1$ we need to subtract the condensed
counterions, as determined by $\rho_1$. We can estimate $\rho_1$
at this point by solving the Poisson-Boltzmann
equation in a unit cell with the
appropriate radius. The chemical potentials of the three ion
species are tuned such that their concentrations match the known
values of $c_z$ and $c_s$. This leads to an estimate:
$c^{\infty}_1 \simeq 68\,{\rm mM}$. Hence, $c_z$ at the onset of
aggregation should lie a little below the continuation of the
$c_s=88\,{\rm mM}$ line which is, indeed, where it is found. The
trend for $c_s = 13\,{\rm mM}$ can probably be seen already at the
point $c_{\rm DNA} = 15\,{\rm mM}$, although the deviation at this
point is still within the range of experimental error. The few
other experimental points with $c_{\rm DNA} \approx c_s$ deviate
slightly from the straight line as well (still within experimental
error bars). In all these cases the deviation is in the direction
corresponding to a higher value of $c_s$, as expected.

A linear relation of the form $c_{z,{\rm aggr}} = \alpha c_{\rm DNA} +
\beta$, was previously suggested on empirical basis for
aggregation induced by spermidine ($3^+$), on a smaller
range of DNA concentrations \cite{OslandKleppe,Pelta}.
Although this result looks similar to our
prediction on the onset of aggregation, it is not directly related
to our analysis because $c_{z,{\rm aggr}}$ was taken in those
works to be the transition midpoint. This is the point where half
of the maximal precipitation of DNA is reached. Our analysis does
not apply at the transition midpoint since it requires all the DNA
segments to be well separated from each other. Indeed, the
coefficient $\alpha$, related to the transition midpoint, was
found in \textcite{OslandKleppe} and \textcite{Pelta} to be of
order $10^2$,  much larger than unity. Such a value of $\alpha$
cannot be interpreted as the excess of spermidine ions per monomer
near isolated chains.

The parameters of the linear fit in Fig.~5 are summarized in
Table~1 for the four experimentally used values of $c_s$.

\begin{table}
\noindent {\small \textbf{TABLE 1 \ \  Fit parameters used in Fig.~5.}}\\
\begin{tabular*}{\linewidth}{l@{\extracolsep{\fill}}ll}
\hline
$c_s$[mM] & $c_z^*$[mM] & $a \rho_z^*$\\
\hline
$2$  & \hspace{\fill} $\ \ \ \ \, 0 \pm 0.0003$ \hspace{\fill} & $0.194 \pm 0.020$ \\
$13$ & $0.011 \pm 0.002$ & $0.191 \pm 0.013$ \\
$23$ & $0.031 \pm 0.005$ & $0.173 \pm 0.025$ \\
$88$ & \ \,$0.52 \pm 0.05$ & $0.135 \pm 0.026$ \\
\hline
\end{tabular*}
\end{table}

\subsection*{Crossover in the log-log plot}

For presentation purposes we plot in Fig~5\,\textit{b}, $c_{z,\rm
aggr}$ vs. $c_{\rm DNA}$ on a log-log scale, as appeared in
\textcite{Raspaud}. The linear relation that was found between
these two quantities is not clearly manifested on the log-log
plot, because a linear dependence of the form $y=\alpha x + \beta$
is not easily recognized in such a plot. Furthermore, such a
linear relation appears on a log-log plot to be
\textit{artificially} characterized by two distinct behaviors, at
low and high values of the independent variable.
These two behaviors were mentioned in
\textcite{Raspaud} and can be seen in Fig.~5\,\textit{b}. However,
they do not represent in our opinion
two real physical regimes and can be
understood by taking the logarithm of Eq.~\ref{onset}.
For small $c_{\rm DNA}$ (large $R$):
\begin{equation}
\log c_z \simeq \log c_z^*
\end{equation}
i.e, $c_z$ does not depend on $c_{\rm DNA}$ as is seen in
Fig.~5\,\textit{b} in the small $c_{\rm DNA}$ limit.
In the opposite limit of large $c_{\rm DNA}$ (small $R$):
\begin{equation}
\log c_z \simeq \log c_{\rm DNA} + \log a\rho_z^*
\end{equation}
Here, the linear dependence of $c_z$ on $c_{\rm DNA}$ yields a
line of slope 1 in the same figure.

The crossover between these apparent behaviors occurs when the
number of bulk and excess ions are the same:
\begin{equation}
c_{\rm DNA} = \frac{c_z^*}{a\rho_z^*} \label{crossover}
\end{equation}
When $c_{\rm DNA}$ is much smaller than this crossover value, the
number of excess multivalent ions near DNA segments is negligible
compared to their total number. In the other extreme of $c_{\rm
DNA}$ much larger than the crossover value, the number of free
multivalent ions is negligible compared to the excess ions, and
nearly all multivalent ions are bound to the DNA.

For the experimental data in Fig.~5 the crossover value is
equal to $0.06$, $0.18$ and $3.9\,{\rm mM}$ for $c_s=13$,
$23$ and $88\,{\rm mM}$, respectively, and smaller than $1.5\times
10^{-3}\,{\rm mM}$ for $c_s=2\,{\rm mM}$. The first
three crossover points are indicated by arrows in Fig.~5\,\textit{b}.


\section*{DNA AGGREGATION AND COUNTERION CONDENSATION}

We separate the discussion following our results in three parts. The
first addresses the conditions required for DNA aggregation. The
coefficients of the linear relation in Eq.~\ref{onset}, $c_z^*$
and $\rho_z^*$, have a definite physical meaning. Their values, as
extracted from the experimental data provide
insight on these conditions.
The second part deals with condensation of counterions on DNA (to
be distinguished from condensation of DNA chains). The general
relation $\rho_z=\rho_z(c_1^\infty,c_z^\infty)$ that was
introduced in Eqs. \ref{totalz} - \ref{excess} is a property of
counterion condensation  on isolated chains. By extracting the
values of $\rho_z$, $c_1^\infty$ and $c_z^\infty$ at the onset of
DNA aggregation, we can learn about exact density profiles of
spermine around DNA, and compare our findings with approximations
such as Poisson-Boltzmann theory.
Finally, we comment on our main assumption, which
was used in the theoretical considerations section.

\subsection*{Conditions at the onset of aggregation}

Most of the proposed theoretical models for 
inter-chain attraction and aggregation (see, for example,
\textcite{delaCruz,Raspaud,Itamar,ItamarPRL,Joanny,Shklovskii,HaLiu,Levin})
regard the charged chain as surrounded by a layer of condensed
ions which is usually modeled as a one-dimensional gas. This
layer mediates an inter-chain attraction, and the models predict
the number of condensed ions required to initiate aggregation of
the chains. In the current work we do not address this
theoretical problem, but rather concentrate on what can be
inferred from the experimental results using the analysis
presented in the previous section. This analysis provides
insight on the conditions prevailing at the
onset of aggregation. In particular, the excess $\rho_z^*$
characterizes the number of condensed multivalent counterions
that are present near each chain at the onset. Although in
general the notion of condensed counterions is
somewhat ill-defined, as it depends on which ions are regarded
as bound to the DNA, we show in the Appendix that in our case it
does have a reasonably well defined meaning. Furthermore, the
number of condensed multivalent ions per monomer
is practically the same as $a\rho_z^*$.

\begin{figure}
\scalebox{0.45}{\includegraphics{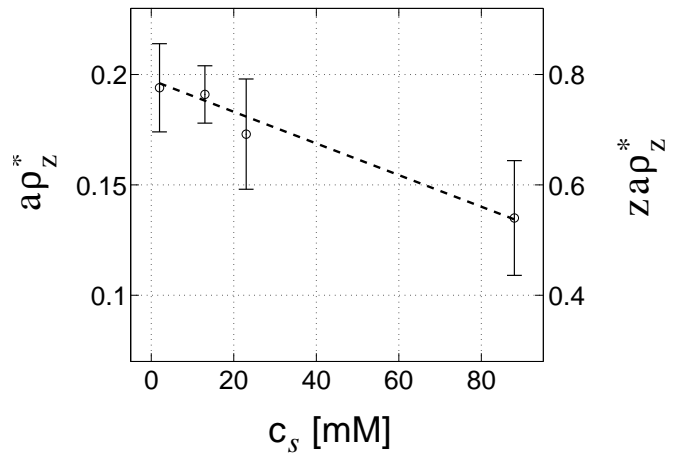}}
\caption{Excess of multivalent counterions per monomer at the
onset of aggregation, $a \rho_z^*$, as function of $c_s$.
All values are taken from Table~1, as extracted
from the experimental data of \textcite{Raspaud}.
Error bars
are indicated by vertical bars and
the dashed line is a linear fit to be used as a guide to the eye.
On the right axis $z a \rho_z^*$ is shown, where $z=4$
for spermine. This value is equal to
the fraction of DNA charge
compensated by the condensed multivalent ions. Note that
according to the Manning condensation theory the same quantity
is equal to 0.94, for tetravalent ions and
no added salt.
}
\end{figure}

The excess of multivalent counterions per monomer, $a \rho_z^*$,
is shown in Fig.~6 as function
of $c_s$. All values are taken from Table~1, as extracted
from the experimental data. The dashed line is a linear fit.
Two different axis scales are used on the left and right of
the plot. The left axis shows the value of $a \rho_z^*$. The
right one shows the part of DNA charge that is compensated
by condensed multivalent ions, $z a \rho_z^*$,
where $z=4$ for spermine. From the plot we deduce the following
two conclusions:

\begin{enumerate}
\item The number of condensed multivalent ions (per DNA monomer)
$a\rho_z^*$ at the onset
decreases as the monovalent salt concentration
increases, with variation between $0.19$ and $0.14$.
A possible reason for this trend may be that
the bare electrostatic repulsion between chains is decreased due to
increased screening. Hence a smaller number of multivalent ions
is required in order to overcome this repulsion.
The change in $\rho_z^*$ may also be related
to the competition between monovalent and multivalent ions in the
aggregated DNA state.

\item The data indicates that there is no over-charging of the DNA
by spermine at the onset (see also \textcite{Shklovskii}) 
since $z a \rho_z^* < 1$. 
At higher concentration of spermine, beyond the threshold, we do not 
rule out the possibility of DNA
over-charging, as was suggested by \textcite{Shklovskii}.
\end{enumerate}

Although $\rho_z^*$ decreases with increase of $c_s$, it
is of the same order of magnitude for all the $c_s$
values in Table~1. In contrast, $c_z^*$ varies in Table~1
over more than three orders of magnitude.
As was previously suggested \cite{delaCruz,Raspaud},
this large variation in $c_z^*$ is a result
of competition between monovalent and multivalent counterions.
We discuss the relation between $\rho_z^*$ and $c_z^*$ to
some extent in the following subsection. A more detailed analysis
of this relation, emphasizing the role of competition between
the two counterion species, will be
presented in a separate publication 
(see also, \textcite{BelloniDrifford,Wilson1,Wilson3}).

\subsection*{Counterion condensation}

We now turn to
analyze the condensation of monovalent and multivalent ions
around DNA. Each line in Table~1 provides a measurement of the
excess $\rho_z$ at certain values of $c^{\infty}_1$ and
$c^{\infty}_z$. The general relation
$\rho_z(c_1^\infty,c_z^\infty)$ is a property of counterion
density profiles around isolated DNA segments. Hence, the data in
Table~1 can be used to test any particular theory used to
calculate such ion distributions.

\begin{table}
\noindent {\small \textbf{TABLE 2 \ \ Excess of 4-valent ions near DNA
compared with PB theory.}}\\
\begin{tabular*}{\linewidth}{l@{\extracolsep{\fill}}llll}
\hline
$c_1^\infty$[mM] & $c^{\infty}_z$[mM] & $a \rho_z$ (exp) & $a\rho_z$ (PB) \\
\hline
$2$  & $\ \ \ \ \ 0 \pm 0.0003$    & $0.194 \pm 0.020$ & $0.186 \pm 0.005$ \\
$13$ & $0.011 \pm 0.002$ & $0.191 \pm 0.013$ & $0.178 \pm 0.002$ \\
$23$ & $0.031 \pm 0.005$ & $0.173 \pm 0.025$ & $0.172 \pm 0.002$ \\
$88$ & $\ \,0.52 \pm 0.05$   & $0.135 \pm 0.026$ & $0.164 \pm 0.002$ \\
\hline
\end{tabular*}
\end{table}

The most simple model to consider is the Poisson-Boltzmann (PB)
theory (see \textcite{Andelman_review,Oosawa,LeBretZimm,Gueron}).
In Table~2 we compare the excess predicted by PB theory
with the experimental result, by solving the PB equation
such that $c_1^\infty$ and $c_z^\infty$ match the experimental
values of $c_s$ and $c_z^*$ from Table~1. The excess
is then calculated from the PB density profile,
and compared with the experimental value of
$a\rho_z$ (equal to $a \rho_z^*$ of Table~1). The DNA is
modeled as a uniformly charged cylinder of radius
$d=10\,\mbox{\AA}$.

Inspection of the results in Table~2 shows that
there is a reasonable agreement with experiment (within the error
bars) for the three smaller values of $c_s$ = 2, 13, 23\,mM.
However, for $c_s =
88\,{\rm mM}$ there is a $30\%$ deviation. The two data points
with $c_{\rm DNA} > 10\,{\rm mM}$ that were not taken into account
in the linear fit of Fig.~5 suggest that $\rho_z$ is
closer to the lower
bound of the experimental error range, whereas the PB value is
larger than the upper bound.

Overall, the agreement with PB theory (Table~2) is surprisingly
good considering that PB theory does not work so well for bulky
multivalent ions. Deviations from PB theory have several sources.
One of these sources is specific molecular details such as the
geometrical shape of ions, DNA structure and short--range
interactions.
Another source for deviations are ion-ion correlations between
spermine molecules, computed in theories which go beyond the mean-field 
approximation. However, these correlations tend to increase
the number of bound multivalent counterions \cite{Lyubartsev3},
while for $c_s=88\,{\rm mM}$, the number of bound multivalent
counterions is decreased. We conclude that correlation effects by
themselves are not the main source of the deviations seen in
Table~2. In addition the data analysis does not indicate
over-charging of the DNA. Such an effect may be expected if
correlation effects are strong \cite{Shklovskii}.

\begin{figure*}
\scalebox{0.45}{\includegraphics{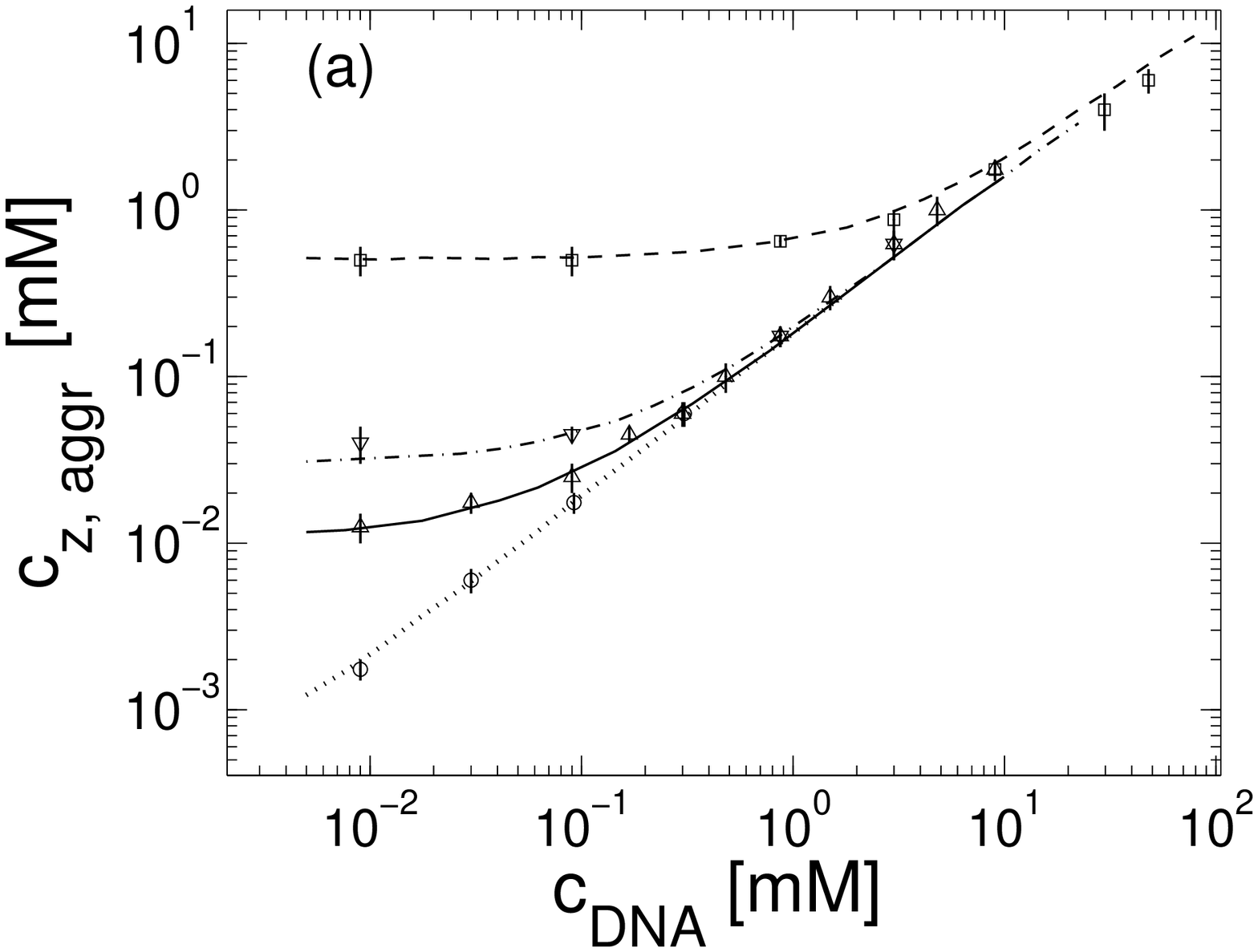}}
\hspace{0.5cm}
\scalebox{0.45}{\includegraphics{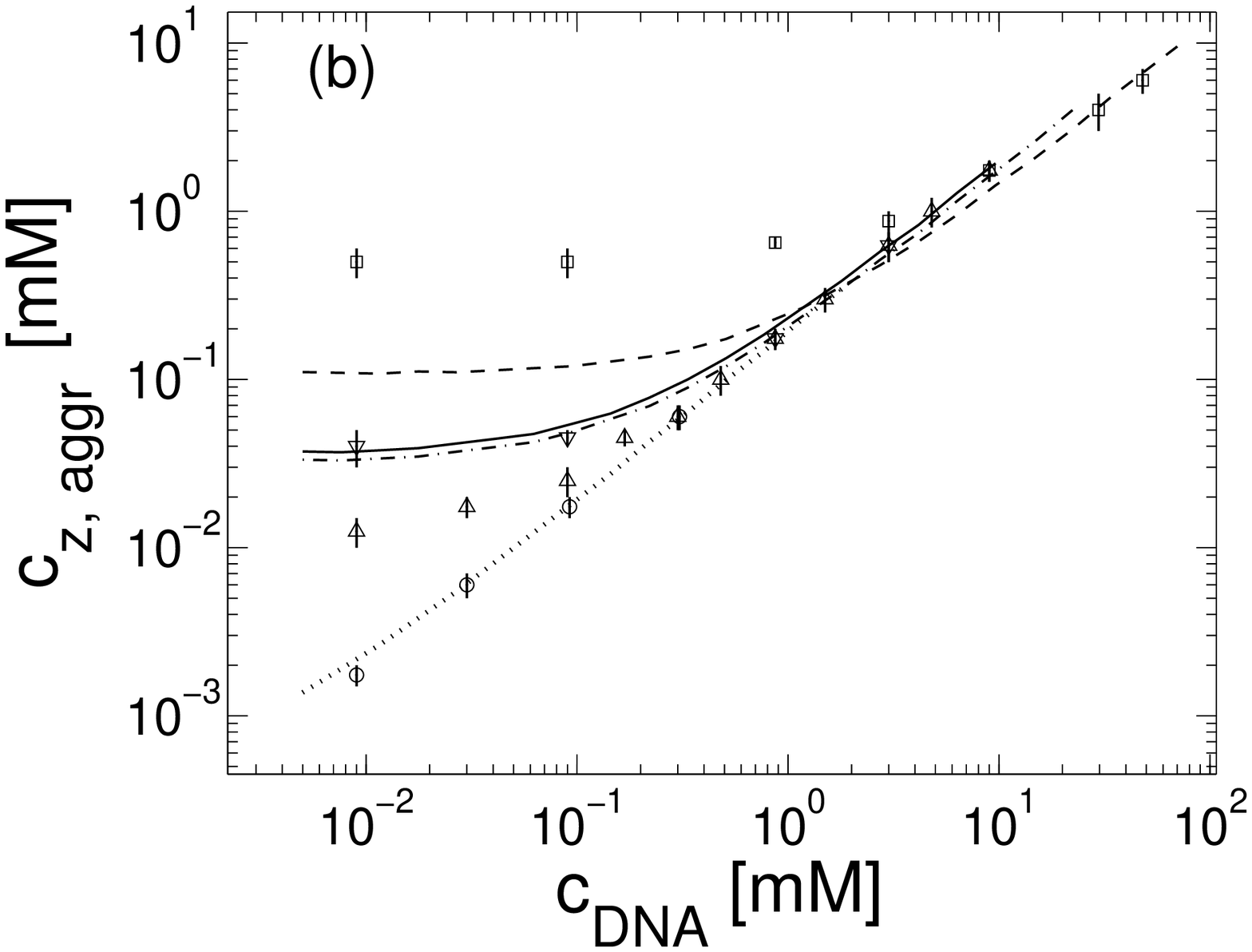}}
\caption{Spermine concentration (in mM) as a
function of DNA monomer concentration (mM) at the onset of
aggregation, calculated using the PB equation. Two different
criteria are used in parts (a) and (b) to determine the onset: in
(a) $c^{\infty}_z$, as calculated using the PB equation, is equal
to the experimental value of $c_z^*$ from Table~1. In (b) $\rho_z$
of PB theory is equal to $\rho_z^*$ from Table~1. The radius of
DNA is taken as $d=10\,\mbox{\AA}$. Log-log plot is used in order
to show the five decades of DNA concentrations. For each $c_s$
the plot covers experimental data up to $c_{\rm DNA} =
c_s$. For larger $c_{\rm DNA}$, corrections due to changes in
$c^{\infty}_1$ should be taken into account, as was discussed in the
preceding section. All notations are the same as in Fig.~5.
}
\end{figure*}

In Fig.~7 we compare the DNA aggregation data with PB predictions
at finite DNA concentrations. For each DNA concentration
the PB equation is solved in a cylindrical cell
of the appropriate radius. The multivalent
counterion concentration $c_z$ is gradually increased until
the onset is reached, and its onset value, $c_{z,{\rm aggr}}$
is plotted as function of $c_{\rm DNA}$. Two
different criteria are used to determine the onset
$c_{z, {\rm aggr}}$.
In Fig.~7\,\textit{a} it is chosen as the point where
$c_z^\infty$ is equal to the experimental value
$c_z^*$ of Table~1; whereas in Fig.~7\,\textit{b} the onset
is chosen the point where $\rho_z=\rho_z^*$.
In order to span all the data range we use for
convenience a log-log plot, as in Fig.~5\,\textit{b}.

On a linear scale all the lines in Fig.~7\,\textit{a} and
\textit{b} are straight lines. This fact serves as additional
confirmation of our general analysis in the theoretical
considerations section. In accordance with our analysis,
both $c_z^*$ and $\rho_z$ are constant along each line,
and the slope of each line is equal to $a\rho_z$.
Note that the relation between $c_z^*$ and $\rho_z$ is determined
in Fig.~7 within the PB approximation, while in Fig.~5 both of these
coefficients are related to the actual counterion density profiles
in the experimental system.
The use of the PB equation is the source of deviations from
experimental data in Fig.~7. 

On first inspection the match with experiment in
Fig.~7\,\textit{a} is very good, whereas the match in
Fig.~7\,\textit{b} is not as good. On closer inspection it is seen
that the fit in Fig.~7\,\textit{b} is not good for small values of
$c_{\rm DNA}$, while it is actually better than in Fig.~7\,\textit{a}
for large $c_{\rm DNA}$. With the PB equation it is not possible
to obtain a perfect fit for both small and large $c_{\rm DNA}$
because the values of $c_z$ and $\rho_z$ are not independent.
Fixing $c_z^\infty=c_z^*$ (as in Fig.~7\,a)
sets a value of $\rho_z$ that
is different from $\rho_z^*$; and the opposite happens in Fig.~7\,b.
The fit in Fig.~7\,\textit{a} is quite good
even for large $c_{\rm
DNA}$ because the values of $\rho_z^*$ are of similar order of
magnitude for all four lines.

Deviations as in Fig.~7 are inevitable if any approximations
are used to model the distribution of counterions around DNA. 
Note however
that within such approximate models our 
general theoretical considerations should apply, 
as long as the total number of ions in the system
is counted properly. Such a model that goes beyond
PB was proposed in 
\textcite{Shklovskii2}. Indeed, within this model a linear relationship 
similar to Eq.~\ref{onset} was found.

The experimental results analyzed in this section may be
influenced, to a certain degree, by the fact that there was more
than one type of monovalent counterion in the system. For the
three higher salt concentrations, except for $c_s = 2$\,mM, the solution contained
$10\,{\rm mM}$ of TrisH$^{+}$ ions in addition to Na$^+$ \cite{Raspaud}. For the
largest salt concentration, $88\,{\rm mM}$, where significant
deviations from PB theory are found, this effect is probably
negligible. Another detail regarding the TE buffer is that 
the Tris ions may be only partly ionized. If only 80\% of
Tris is ionized, as suggested in \textcite{TangJanmey}, the
concentrations $c_s = 13$\,mM, 23\,mM and 88\,mM 
should be reduced by 2\,mM. 
Although this will have only a small effect
on our results, it will improve both the comparison with PB
and the fit with the dashed line in Fig.~6, for the
point $c_s = 13$\, mM. For the two other concentrations of 23\,mM and 88\,mM 
the effect will be negligible.

\subsection*{Further comments on underlying model assumption}

Our underlying assumption, that the onset of aggregation
depends uniquely
on $c_1^\infty$ and $c_z^\infty$ (but not on $c_{\rm DNA}$), 
is an approximation that
can be justified on several different levels but deserves further
and more thorough investigation.
The most simple motivation for this assumption is that multivalent 
ions, in the vicinity of the chains,
mediate the attraction necessary for aggregation. 
In turn, the number of condensed multivalent ions 
near each chain is
determined by $c_1^\infty$ and $c_z^\infty$. 

Let us first suppose that aggregation starts when a net
attraction appears between two chains.
This assumption may be justified if chains are
sufficiently long and their 
translational entropy can be neglected. 
In order to find the onset of two-chain attraction
the free energy of a two-chain
complex should be calculated 
as a function of the  distance between the two chains.
This free energy represents the effective interaction
between the two chains, mediated by the ionic solution.
The counterion distribution near each chain
will not be the same for close-by and for isolated
chains. However in both cases the concentrations
must decay to their bulk values
throughout the solution, $c_1^\infty$ and $c_z^\infty$. 
This requirement serves as a boundary condition, 
imposed at a large distance from the two chains. 
It will determine uniquely the counterion distribution 
between the chains, as well as the free energy associated
with the two-chain complex. 
Hence $c_1^\infty$ and $c_z^\infty$ determine the effective
interaction between chains, and in particular
whether an attraction occurs at a certain range of inter-chain
separations; in terms of these variables the onset 
of two-chain attraction does not depend on $c_{\rm DNA}$.

Strictly speaking, the onset of aggregation and the onset of
two-chain attraction are not the same. The aggregate phase
involves interactions between multiple chains, whereas chains in
the dilute phase interact very weakly with each other. Aggregation
starts when the free energy per chain is equal in the dilute and
aggregate phases. 
Note that the chemical potential of each  ion species must be the same
in the two phases, and that in the dilute phase these chemical
potentials are directly related to $c_1^{\infty}$ and
$c_z^{\infty}$. Hence $c_1^\infty$ and $c_z^\infty$ determine the
free energy per chain in the two phases. The approximation of
independence on $c_{\rm DNA}$ neglects the 
translational entropy of DNA
segments, which can be justified for long enough and rigid
segments. It also neglects contributions from interactions between
chains in the dilute phase, which are assumed to be small compared
to the free energy of the single DNA-counterion complexes.

\section*{SUMMARY}

We have shown that the onset of aggregation at finite (non-zero)
DNA concentration, $c_{z,{\rm aggr}}$, is determined by the onset
in the limit of infinite DNA dilution. For DNA monomer concentration smaller than
that of monovalent salt, $c_{\rm DNA} \lesssim c_s$, the
multivalent counterion concentration at the onset,  $c_{z,{\rm aggr}}$, depends linearly
on $c_{\rm DNA}$. The coefficients of this linear dependence are
the bulk concentration of multivalent counterions and their excess
relative to the bulk near each DNA segment. Both of these coefficients
are of theoretical interest and can be extracted from the
available experimental data.

Our main assumption is that the onset of aggregation can be
related to the ion density profiles around each chain. Hence, it is
uniquely determined by $c_1^\infty$ and $c_z^\infty$,
the bulk concentrations of the two counterion species, respectively.
Our results and fit to experiment strongly support
this assumption. Nevertheless, we believe that
more detailed theoretical and experimental investigations
are needed in order to fully understand its range of validity.
For example, it will be of interest to test  experimentally the
equilibration of a DNA solution through a
dialysis membrane, with a cell containing only
counterions \cite{Braunlin,PlumBloomfield,SubiranaVives}.
This procedure allows a direct
control of the ionic bulk concentrations.

In order to predict precisely the onset of aggregation, the
structure of the aggregated phase must be considered.
Nevertheless, it is instructive to focus only on single chains at
the onset, as is often done. At the aggregation onset the
electrostatic repulsion between isolated chains in solution
must be overcome by
a sufficiently strong attraction mediated by multivalent
counterions. This number of counterions is expected to depend only
weakly on physical parameters such as the monovalent salt
concentration. Our analysis does not address directly the question
of the onset origin, but  merely supports the fact that the number
of condensed multivalent ions at the onset, $a\rho_z^*$,
is of the same order of magnitude,
regardless of the $c_s$ value. A more refined result of our analysis
is that $a\rho_z^*$ is not constant but
decreases with increase of $c_s$. On the other hand $c_z^*$, the value of $c_z^\infty$
at the onset,
depends strongly on $c_s$. This is mainly a result of the
competition between monovalent and multivalent ions, as
will be addressed in a separate publication.

Our analysis also sheds light on counterion condensation on DNA,
which is independent on the criterion for DNA aggregation. The
experimental data indicates that for high $c_s$ the number of
spermine ions in the vicinity of DNA is smaller than the
prediction of Poisson-Boltzmann theory. A similar trend was observed in
computer simulations \cite{Lyubartsev3} of spermidine (3$^+$)
and NaCl in contact
with DNA. Spermidine binding was affected by addition of
monovalent salt more strongly than the Poisson-Boltzmann
prediction. For high
salt concentrations spermidine binding was considerably smaller.
In the computer
simulations both molecular specific interactions, the geometrical
shape of the constituents and inter-ion correlations were taken
into account. All these effects, and in particular the geometry of
the spermidine molecule, which is similar to that of spermine,
were found to play an important role.

The above analysis demonstrates that specific interactions 
play an important role in determining the threshold 
of aggregation. In the dilute phase these interactions strongly 
influence the competition between monovalent and multivalent ions
and the free energy of DNA-counterion complexes. 
Similarly, specific interactions play a prominent role 
in the dense phase \cite{Parsegian3}. Force measurements under osmotic
stress \cite{Parsegian0,Parsegian1,Parsegian2}
provide a wealth of information on these interactions.

In conclusion, the physical parameters extracted here from experiment
on the onset of DNA aggregation provide insight on the conditions required
for aggregation, and on condensation of ions around DNA. These
parameters may turn out to be of great value in assessment of
various theoretical models. Additional detailed experiments
may further deepen our understanding of these complex phenomena.

\vspace{2cm}

{\small We are grateful to E. Raspaud and J.-L. Sikorav for
numerous discussions and for providing us with their unpublished
experimental data. We also wish to thank I. Borukhov, 
H. Diamant, M. Kozlov, A. Lyubartsev, G. Manning, T. Nguyen,
M. Olvera de la Cruz, A. Parsegian, R. Podgornik, D. Rau and 
T. Thomas for discussions and 
correspondence. Support from the U.S.-Israel
Binational Science Foundation (B.S.F.) under grant No. 98-00429
and the Israel Science Foundation under grant No. 210/02 is
gratefully acknowledged. One of us (DA) thanks the Alexander von
Humboldt Foundation for a research award.}


\section*{APPENDIX}

In this appendix we discuss the relation
between the excess and the number of condensed ions. The latter
quantity is not as well defined as the former, but relates more
naturally to the aggregation mechanism.
The notion of condensed ions suggests that some ions are bound to
the charged chain while others are free. In reality there is a
density profile that extends all the way from $r=d$ to $r=R$
with no definite separation between condensed and free ions.
In the following we define condensed ions rather loosely as the
number of ions up to a certain characteristic distance from the
chain \cite{BelloniDrifford,Wilson3}. 
We show that for multivalent ions this number does not
depend strongly on the choice of this characteristic distance.
Hence, the number of condensed ions is reasonably well defined.
Moreover, the excess number of multivalent counterions, which can
be directly calculated from the experimental data, is nearly
identical to this quantity. This point will be further explained below.

\begin{figure}
\scalebox{0.42}{\includegraphics{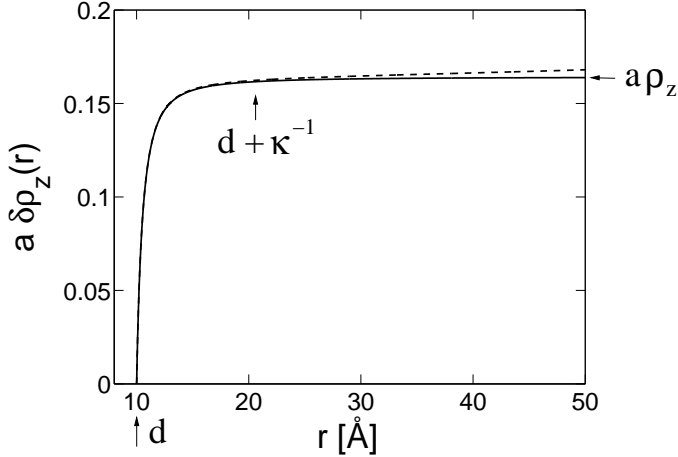}}
\caption{Excess of 4-valent ions per DNA
monomer, up to a distance $r$ from the axis of a charged cylinder
 of radius $d=\,10\mbox{\AA}$ (modeling the DNA)
as obtained using the
Poisson-Boltzmann equation (solid line). 
The excess $\delta \rho_z(r)$ is defined in
Eq.~\ref{excessofr}. The number of charges per unit length on the
cylinder is $1/a$ where $a = 1.7\,\mbox{\AA}$ to fit DNA
values. The bulk densities of monovalent and multivalent ions are
$c_1^{\infty}=88\,{\rm mM}$, $c_z^{\infty}=0.52\,{\rm mM}$, yielding
$\kappa^{-1}=10.0$\,\AA. The quantity $\delta\rho_z$ (solid
line) can be compared with the total number of
4-valent ions (dashed line)
up to a distance $r$ from the cylinder. The distance
$d+\kappa^{-1}$ from the DNA axis is indicated by a vertical
arrow, and characterizes the decay of the density profile far
away from the DNA.
}
\end{figure}

Fig.~8 shows the excess of 4-valent counterions $\delta \rho_z(r)$
up to a distance
$r$ from the DNA axis, as a function of $r$:
\begin{equation}
\tag{A1}
\delta\rho_z(r) = 2\pi \int_0^r r'{\rm
d}r'\,\left[{n_z(r')-c^{\infty}_z}\right] \label{excessofr}
\end{equation}
with the limit $\delta\rho_z(\infty)=\rho_z$ of Eq. \ref{excess}.
The density profile was calculated using the Poisson-Boltzmann
equation, with the
radius of DNA taken as $d=10\,\mbox{\AA}$ and with bulk densities
of ions as in the last line of Table~1: $c_s =c_1^\infty= 88\,{\rm
mM}$, $c_z^{\infty} = 0.52\,{\rm mM}$.

Three observations can be made. First,
most but not all of the excess z-valent ions are localized
very close to the DNA, at a distance of order $\lambda / z$, where
$\lambda$ is the Gouy-Chapman length (see \textcite{Andelman_review}):
\begin{equation}
\tag{A2}
\lambda = \frac{1}{2\pi l_B \sigma} = \frac{d}{l_B \rho_{\rm DNA}}
\end{equation}
where $\sigma$ is the average charge per unit area on the cylinder
surface, $\sigma = \rho_{\rm DNA}/2\pi d$, and $\rho_{\rm DNA}=1/a$
is the DNA charge per unit length. At room temperature the
Bjerrum length $l_B \simeq 7\,\mbox{\AA}$ and for DNA with
4-valent counterions $\lambda/z \simeq 0.6\,\mbox{\AA}$.
Second,
the counterions within a layer of few times the Debye length
($\kappa^{-1} = 10.0\,\mbox{\AA}$ in Fig.~8) neutralize the DNA
charge. Nearly all the excess distribution is in this layer.
Third, in order to
estimate the total amount of counterions in the condensed layer
of thickness $\alpha\kappa^{-1}$, where $\alpha$ is a number of
order unity, we need to add $\delta\rho_z$ to the bulk
contribution, $\pi\alpha^2\kappa^{-2}c_z^\infty$. Using $\kappa$
from Eq. \ref{kappadef}, the latter is equal to:
\begin{equation}
\tag{A3}
\left(\frac{\alpha^2}{4 l_B}\right) \frac{c^{\infty}_z}{2
c^{\infty}_1 + z(z+1)c^{\infty}_z} \label{estimate}
\end{equation}
In experiment, $c_z^{\infty}$ is much smaller than $c_1^{\infty}$
at the onset, and the bulk contribution of Eq. \ref{estimate} can
be neglected relative to $\rho_z$, for $\alpha$ of order unity.
This can be seen specifically in Fig.~8 by comparing the solid and
dashed lines.

The outcome of the above discussion is that $\rho_z$, defined in
Eq. \ref{excess}  as the excess of counterions throughout the
cell, can be regarded, to a good approximation, as the total
number of counterions within a condensation layer whose thickness is
approximately the Debye length. For typical
concentration ranges as considered here we do not expect that this
outcome will change, even for models going beyond
Poisson-Boltzmann theory.

As a further demonstration, the number of
multivalent counterions up to several different distances
from the DNA is shown in Table~3, as calculated in a unit cell
using the Poisson-Boltzmann equation. For each $c_s$ in Table~1 we find
the Poisson-Boltzmann density profile such that $c_1^\infty = c_s$
and $\rho_z = \rho_z^*$, and then calculate the number
of multivalent ions (per DNA monomer)
up to the following distances from the DNA radius:
10\,\AA, 20\,\AA, $\kappa^{-1}$ and $2 \kappa^{-1}$.
The values of $\kappa^{-1}$,
as obtained from Eq.~\ref{kappadef}
are equal to 68, 26, 20 and 10\,\AA~
for $c_s = $ 2, 13, 23 and 88\,mM, respectively.
These numbers are compared with $a \rho_z^*$. All the
different measures in Table~3
yield results that are very close to each other.

\begin{table}
\noindent {\small \textbf{TABLE 3 \ \ Number of z-valent
counterions, per DNA monomer, up to several
different distances from the DNA axis,
compared with $a\rho_z$.}}\\
\begin{tabular*}{\linewidth}{l@{\extracolsep{\fill}}lllll}
\hline
$c_s$[mM] & d + 10\,\AA & d + 20\,\AA
& d + $\kappa^{-1}$ & d + $2\kappa^{-1}$ & $a \rho_z$ \\
\hline
$2$  & 0.191 & 0.193 & 0.194 & 0.194 & 0.194 \\
$13$ & 0.187 & 0.190 & 0.190 & 0.191 & 0.191 \\
$23$ & 0.171 & 0.172 & 0.172 & 0.173 & 0.173 \\
$88$ & 0.134 & 0.135 & 0.134 & 0.135 & 0.135 \\
\hline
\end{tabular*}
\end{table}

\end{document}